\documentclass[prb,floatfix,twocolumn,showpacs,superscriptaddress,preprintnumbers,amsmath,amssymb]{revtex4}
\usepackage{graphicx}
\usepackage{dcolumn}
\usepackage{bm}
\usepackage[dvips]{color}

\usepackage{graphpap}

\begin{document}

\title{Engineering Silicon Nanocrystals: Theoretical study of the effect of Codoping with Boron and Phosphorus}

\author{Federico \surname{Iori}}
\affiliation{CNISM-CNR  and Dipartimento di Fisica, Universit\`a di Modena e Reggio Emilia, via Campi 213/A, I-41100
Modena, Italy}
\author{Elena \surname{Degoli}}
\affiliation{CNR-INFM-$S{^3}$ and Dipartimento di Scienze e Metodi dell'Ingegneria, Universit\`a di Modena e Reggio Emilia, via Amendola 2 Padiglione Morselli, I-42100 Reggio Emilia, Italy}
\author{Rita \surname{Magri}}
\affiliation{CNISM-CNR  and Dipartimento di Fisica, Universit\`a di Modena e Reggio Emilia, via Campi 213/A, I-41100 Modena, Italy}
\author{Ivan \surname{Marri}}
\affiliation{CNR-INFM-$S{^3}$ and Dipartimento di Scienze e Metodi dell'Ingegneria, Universit\`a di Modena e Reggio Emilia, via Amendola 2 Padiglione Morselli, I-42100 Reggio Emilia, Italy}
\author{G. \surname{Cantele}}
\affiliation{CNR-INFM-Coherentia and Universit\`a di Napoli ``Federico II'', Dipartimento di Scienze Fisiche, Complesso Universitario Monte S. Angelo, Via Cintia, I-80126 Napoli, Italy}
\author{ D. \surname{Ninno}}
\affiliation{CNR-INFM-Coherentia and Universit\`a di Napoli ``Federico II'', Dipartimento di Scienze Fisiche, Complesso Universitario Monte S. Angelo, Via Cintia, I-80126 Napoli, Italy}
\author{F. \surname{Trani}}
\affiliation{CNR-INFM-Coherentia and Universit\`a di Napoli ``Federico II'', Dipartimento di Scienze Fisiche, Complesso Universitario Monte S. Angelo, Via Cintia, I-80126 Napoli, Italy}
\author{O. \surname{Pulci}}
\affiliation{European Theoretical Spectroscopy Facility (ETSF) and
CNR-INFM, Dipartimento di Fisica, Universit\`a di Roma ``Tor
Vergata'', Via della Ricerca Scientifica 1, I-00133 Roma, Italy}
\author{Stefano \surname{Ossicini}}
\affiliation{CNR-INFM-$S{^3}$ and Dipartimento di Scienze e Metodi dell'Ingegneria, Universit\`a di Modena e Reggio Emilia, via Amendola 2 Padiglione Morselli, I-42100 Reggio Emilia, Italy}

\begin{abstract}
We show that the optical and electronic properties of
nanocrystalline silicon can be efficiently tuned using impurity
doping. In particular, we give evidence, by means of ab-initio
calculations, that by properly controlling the doping with either
one or two atomic species, a significant modification of both the
absorption and the emission of light can be achieved. We have
considered impurities, either boron or phosphorous (doping) or
both (codoping), located at different substitutional sites of
silicon nanocrystals with size ranging from 1.1 nm to 1.8 nm in
diameter. We have found that the codoped nanocrystals have the
lowest impurity formation energies when the two impurities occupy
nearest neighbor sites near the surface. In addition, such systems
present band-edge states localized on the impurities giving rise
to a red-shift of the absorption thresholds with respect to that
of undoped nanocrystals. Our detailed theoretical analysis shows
that the creation of an electron-hole pair due to light absorption
determines a geometry distortion that in turn results in a Stokes
shift between adsorption and emission spectra. In order to give a
deeper insight in this effect, in one case we have calculated the
absorption and emission spectra going beyond the single-particle
approach showing the important role played by many-body effects.
The entire set of results we have collected in this work give a
strong indication that with the doping it is possible to tune the
optical properties of silicon nanocrystals.
\end{abstract}
\maketitle
\newpage
\section{Introduction}
Bulk silicon is an indirect band-gap material emitting in the
infrared region. Radiative lifetimes of excited carrier
are very long, causing a predominant de-excitation via
fast non-radiative recombinations. Moreover, silicon has a
significant free carrier absorption and Auger recombination rates
which makes the use of this material in optoelectronic applications very problematic.

During the last decade, several breakthroughs have increased the
hopes of using nanostructured silicon as an optical active
material.\cite{ossobook,ossobori} The basic idea has been to take
advantage of the reduced dimensionality of the nanocrystalline
phase (1-5 nm in size) where quantum confinement, band folding and
surface effects play a crucial role.\cite{ossobook,bisi} Indeed,
it has been found that Si nanocrystals band-gap increases with
decreasing size with a luminescence external efficiency in excess
of 23\%.\cite{ossobook,bisi,gelloz} Moreover, optical gain has
been already demonstrated in a large variety of experimental
conditions.\cite{pav1,pav2,pav3,pav4,lute} Nevertheless, Si
nanocrystals (Si-nc) still have a memory of the indirect band gap
of the bulk phase and this is evidenced by the clearly observed
structures related to momentum-conserving
phonons.\cite{ossobook,canh,del2} This drawback can be
circumvented by introducing an isoelectronic impurities.
\cite{ossobook,bisi} Indeed, in a series of recent very
interesting papers, Fujii and collaborators
\cite{fujii3,fujii1,fujii2} have proved the possibility of a
detailed control of the Si-nc photoluminescence by the
simultaneous doping with {\it n}- and {\it p}- type impurities. In
particular, they have showed that a (B and P) codoped Si-nc always
has an higher photoluminescence intensity than that of both a
single (B or P) doped and of an undoped nanocrystal. Besides,
under resonant excitation condition, the codoped samples did not
exhibit structures related to momentum-conserving phonons
suggesting that, in this case, the quasi-direct optical
transitions are predominant.\cite{fujii3,fujii1,fujii2}

From a theoretical point of view, investigations of impurities in
silicon nanostructures are very few when compared to the large
number of papers reporting calculations for pure, undoped systems;
moreover, most of them are based on semi-empirical approaches. An
handful number of first-principles studies has been devoted to
quantum confinement effects in single-doped
Si-nc.\cite{melnikov,cantele,zhou,ferna} These works have
basically shown that i) the Si-nc ionization energy is virtually
size independent, ii) the impurity formation energy is greater for
smaller nanocrystals and iii) impurity segregation strongly
affects the conductance properties of
nanostructures.\cite{melnikov,cantele,zhou,ferna}

We have recently started a systematic study of the electronic and
optical properties of codoped Si-nc. Our first results
\cite{feffe} show that codoped Si-nc undergo a minor structural
distortion around the impurities and that the formation energies
are always smaller than those of the corresponding single-doped
cases. Moreover, we have found that codoping reduces the band-gap
with respect to the undoped nanocrystals, showing the possibility
of an impurity based engineering of the Si-nc optical
properties.\cite{feffe,ossofeffe}

We report here a comprehensive theoretical study of the
structural, electronic and optical properties of B and P
simultaneously doped Si nanocrystals using {\em ab initio} Density
Functional Theory. The paper is organized as follows. Section
\ref{comp} is focused on the description of the theoretical and
computational methods, section \ref{res} is dedicated to the
discussion of our results. The results will be presented and
discussed in the following order: i) structural properties
(subsection \ref{geometri}), ii) formation energies (subsection
\ref{forme}) and iii) electronic (subsection \ref{elpr}) and iv)
optical properties (subsection \ref{opti}) for all the considered
Si nanocrystals. Concerning the optical properties we discuss both
absorption and emission spectra obtained within a single-particle
approach (\ref{abs}) and, in one case, with many-body methods
considering self-energy corrections and the Bethe-Salpeter scheme
(\ref{lum}). Finally, in section \ref{conc} we summarize our
results.

\section{Computational Methods}
\label{comp} Our results are obtained within a plane-wave
pseudopotential DFT scheme, using the Quantum-ESPRESSO package
\cite{pwscf}. The impurity states are calculated in an
approximately spherical Si-nc, built by considering all the bulk
Si atoms contained in a sphere (centered on a Si ion) with a
diameter ranging from 1.1 nm (Si$_{35}$H$_{36}$) to 1.79 nm
(Si$_{147}$H$_{100}$). The surface dangling bonds are saturated
with hydrogens. Following the work of Fujii et al. \cite{fujii3},
we have located the B and P impurities in substitutional positions
just below the nanocrystal surface. It is worth mentioning that
this arrangement represents the most stable configuration, as
confirmed by theoretical and experimental
works.\cite{feffe,colombi,Garrone} Full relaxation with respect to
the atomic positions has been allowed for both doped and undoped
systems. All the DFT calculations are performed within the
generalized gradient approximation (GGA) using Vanderbilt
ultrasoft \cite{vanderbilt} pseudopotentials for the determination
of both the structural and electronic properties (see Sec.
\ref{geometri}, Sec. \ref{forme} and Sec. \ref{elpr}) whereas
norm-conserving pseudopotential within the Local Density
Approximation (LDA) at the relaxed geometry have been used to
evaluate the optical properties (see Sec. \ref{opti}). This choice
is due to the fact that although Vanderbilt ultrasoft
pseudopotentials allow the treatment of several hundreds of atoms
per unit cell in the atomic relaxation process, the removal of the
norm-conservation condition is a well known problem for the
calculation of the optical transition matrix elements
\cite{check}. Each nanocrystal has been placed in a large
supercell in order to prevent interactions between the periodic
replicas (about 6 {\AA} of vacuum separates neighbor
nanocrystals). Structural, electronic and optical properties, as
well as the impurity formation energies, are investigated as a
function of the size and for several impurity positions within the
Si-nc. Starting from the Si$_n$H$_m$ nanocrystal,\cite{degoli} the
formation energy of the neutral B or/and P impurities can be
defined as the energy needed to insert one B or/and one P atom
within the nanocrystal after removing one/two Si atoms
(transferred to the chemical reservoir, assumed to be bulk Si)\cite{Zhang}\\
\begin{eqnarray}
E_f = E\,( \text{Si}_{n-l-k} \text{B}_{k} \text{P}_{l}
\text{H}_m ) - E\,( \text{Si}_n \text{H}_m )\nonumber\\
+ (k+l)\,\mu_{\text{Si}} - k\,\mu_{\text{B}} - l\,\mu_{\text{P}},
\label{DeltaHf}
\end{eqnarray}
where E is the total energy of the system, $\mu_{\text{Si}}$ the
total energy per atom of bulk Si and $\mu_{\text{B(P)}}$ the total
energy per atom of the impurity (we consider the total energy per
atom in the tetragonal B$_{50}$ crystal for B, as in Ref.
\onlinecite{muB} and the orthorhombic black phosphorus for P, as
in Ref. \onlinecite{muP}). The integers $k$ and $l$ can be set to
either $0$ or $1$. In particular, $k=1$ is the choice  when a B
impurity is present in the nanocrystal ($0$ otherwise) and $l=1$
for a P impurity ($0$ otherwise). With this prescription,
Eq.\ref{DeltaHf} can be used for both the single doping and the
codoping case.

The calculations of the optical properties have been done both in
the ground and in the first excited states, where the excited
state corresponds to the electronic configuration in which the
highest occupied single-particle state (HOMO) contains a hole,
while the lowest unoccupied single-particle state (LUMO) contains
the corresponding electron.\cite{franceschetti,puzder,onida,leo}
It is worth pointing out that an undoped and relaxed Si-nc have
T$_d$ symmetry; in the presence of doping, this high symmetry is
generally lost due to the presence of the impurity atoms.
Moreover, because of the significant differences in the charge
density of the ground and excited states, the actual atomic
relaxations in the two cases are different.

The nanocrystal optical response is evaluated for both the ground
and the excited state relaxed geometries computing the imaginary
part of the dielectric function ($\epsilon_2(\omega)$) through the
Fermi golden rule. The emission spectra have been calculated using
the excited state atomic positions and the ground state electronic
configuration (more details can be found in section \ref{lum}). It
should be noted that although $\epsilon_2(\omega)$ should only be
used for calculating the nanocrystal absorption coefficient, it
can also be used for getting a first approximation to the emission
spectra simply because the emission can be viewed as the time
reversal of the absorption.\cite{bassani} In other words, once the
relaxed atomic positions corresponding to to an hole in the HOMO
and an electron in the LUMO have been found, these atomic
positions are used for the calculation of $\epsilon_2(\omega)$
whose main features are also those of the emission spectra. It is
worth mentioning that the photoluminescence spectra can be derived
using the well known Van Roosbroeck and Shockley \cite{shockley}
relation which, again, involves $\epsilon_2(\omega)$. However,
such a calculation requires the knowledge of the electron and hole
populations, at the working temperature, in the LUMO and HOMO
states respectively. The populations, in turn, depends on the
actual dynamics in the excitation and emission processes,
including the non radiative electron hole recombinations. In this
work we have not considered any particular dynamics so that our
emission spectra contains only the informations related to both
the transition energies and the oscillator strengths.
\begin{table}[!h]
\caption{\label{tableSi87} Bond lengths (in {\AA}) around the
impurity sites for the undoped, single doped and codoped
Si$_{87}$H$_{76}$ nanocrystal (diameter 1.50 nm). B and P
impurities have been substitutionally located at subsurface
positions (see Fig. \ref{si85bp}). Si$_s$ and Si$_i$ refer to two
surface and two inner Si atoms around this site respectively.}
\begin{ruledtabular}
 \begin{tabular}{lccccc}
 & {Si$_{87}$H$_{76}$} & {}&{Si$_{86}$BH$_{76}$} & {Si$_{86}$PH$_{76}$} & {Si$_{85}$BPH$_{76}$} \\
{Bond} & {\AA} & Bond & {\AA} & {\AA} & {\AA} \\
\hline
Si-Si$_s$ & 2.355   &  B-Si$_s$     & 2.036     & & 2.021  \\
Si-Si$_s$ & 2.355   &  B-Si$_s$     & 2.036     & & 2.021\\
Si-Si$_i$ & 2.363   &  B-Si$_i$  &  2.014    & & 2.034\\
Si-Si$_i$ &  2.363  &  B-Si$_i$     &  2.014    & & 2.034\\
\hline
Si-Si$_s$ & 2.355   &  P-Si$_s$  &   &   2.294   & 2.295\\
Si-Si$_s$ & 2.355   &  P-Si$_s$  &   &   2.294   & 2.295\\
Si-Si$_i$ & 2.363   &  P-Si$_i$     &   &   2.380   & 2.331\\
Si-Si$_i$ & 2.363   &  P-Si$_i$     &   &   2.380   & 2.331\\
\end{tabular}
\end{ruledtabular}
\end{table}
In the case of the Si$_{33}$BPH$_{36}$ codoped Si-nc, going beyond
the single-particle approach, we have included the self-energy
corrections by means of the GW approximation.\cite{hedin} In a
successive step, excitonic effects are include solving the
Bethe-Salpeter equation.\cite{onida} A further advantage of this
procedure is that the inhomogeneity of the system is taken into
account by properly including local fields effects.\cite{tra1}
This approach, in which many-body effects are combined with a
study of the structural distortion due to the impurity atoms in
the excited state, allows a precise determination of the Stokes
shift between absorption and emission spectra.\cite{leo}

\section{Results}
\label{res} This section collects all the results we have obtained
in the study of the structural, electronic and optical properties
of boron and phosphorus codoped silicon nanocrystals. When
possible our outcomes are compared with available experimental
results.

\subsection{Structural Properties}
\label{geometri}
First of all, it is interesting to look at the change in the
nanocrystal structure induced by the presence of the impurities.
As outlined above, the B or/and P impurity atoms have always been
located in substitutional sites in the Si shell just below the
surface, these positions having previously been shown to be the
most stable ones.\cite{cantele} Initially, we have considered impurities
located on opposite sides of the nanocrystals, thus at the largest
possible distance.

\begin{figure}
\includegraphics[clip,width=0.35\textwidth]{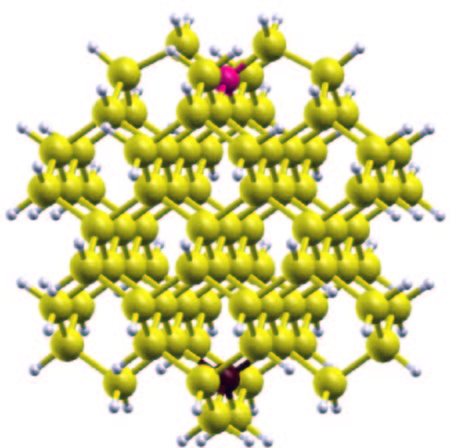}
\caption{\label{si85bp}(Color online) Relaxed structure of
Si$_{85}$BPH$_{76}$ (d = 1.50 nm). Yellow (grey) balls represent
Si atoms, while the white (light grey) balls are the hydrogens
used to saturate the dangling bonds. B (magenta, dark grey) and P
(black) impurities have been located at subsurface position in
substitutional sites on the nanocrystal opposite side. The relaxed
impurity-impurity distance is D$_{\text{BP}}$ = 10.60 {\AA}.}
\end{figure}

Table I gives the relaxed bond lengths around the impurities for
Si$_{87}$H$_{76}$ whose structure is shown in Fig.\ref{si85bp}.
Comparing these bond lengths with those of the corresponding Si
atoms in the undoped Si-nc it is clear that some significant
relaxation occurs around the impurities. In all the cases the
local structure has a $C_{2v}$ symmetry, with two shorter and two
longer Si-impurity bonds with respect to the two surface and the
two inner Si atoms. An interesting point is that the amount of
relaxation around the impurity is directly related to the impurity
valence. The most significant relaxation is found for the
trivalent atom  (B, 2.036 and 2.014 \AA ~with respect to 2.355 and
2.363 \AA) to be compared with that of the pentavalent one (P,
2.294 and 2.380 \AA ~with respect to 2.355 and 2.363 \AA).
Besides, it is interesting to note that in the codoped case the
differences among the four impurity-Si bond lengths are always
smaller than the single-doped case (the Si-B bonds differ of about
1.08\% in the single-doped case and only 0.64\% in the codoped
case, whereas this variation in the case of P reduces from 3.61\%
to 1.54\%). Thus, if carriers in the Si-nc are perfectly
compensated by simultaneous  {\it n}- and {\it p}-type doping, an
almost $T_{d}$ configuration is recovered.

\begin{table}[!h]
\caption{\label{tableSi35}Bond lengths (in {\AA}) around the
impurity sites for the undoped, doped and codoped
Si$_{35}$H$_{36}$ nanocrystal (d = 1.10 nm). Substitutional B and
P impurities are located at subsurface positions (see Fig.
\ref{Si33BP}). Si$_s$ and Si$_i$ label, respectively, the three
surface and one inner Si atoms around this site.}
\begin{ruledtabular}
\begin{tabular}{lccccc}
& {Si$_{35}$H$_{36}$} & {}&{Si$_{34}$BH$_{36}$} & {Si$_{34}$PH$_{36}$} & {Si$_{33}$BPH$_{36}$} \\
{Bond} & {\AA} & Bond & {\AA} & {\AA} & {\AA}\\
\hline
Si-Si$_s$ & 2.300   &   B-Si$_s$     & 2.093     & & 2.035  \\
Si-Si$_s$ & 2.300    &  B-Si$_s$     & 2.022     & & 2.026\\
Si-Si$_s$ & 2.300    &  B-Si$_s$     & 2.022    & & 2.026\\
Si-Si$_i$ & 2.361   &  B-Si$_i$     & 2.008   & & 2.007\\
\hline
Si-Si$_s$ & 2.300   &  P-Si$_s$     &   &   2.366  & 2.303\\
Si-Si$_s$ & 2.300  &   P-Si$_s$     &   &   2.365  & 2.302\\
Si-Si$_s$ & 2.300   &  P-Si$_s$     &   &   2.364   & 2.297\\
Si-Si$_i$ & 2.361   &  P-Si$_i$     &   &   2.310   & 2.334\\
\end{tabular}
\end{ruledtabular}
\end{table}

\begin{figure}[!b]
\includegraphics[clip,width=0.35\textwidth]{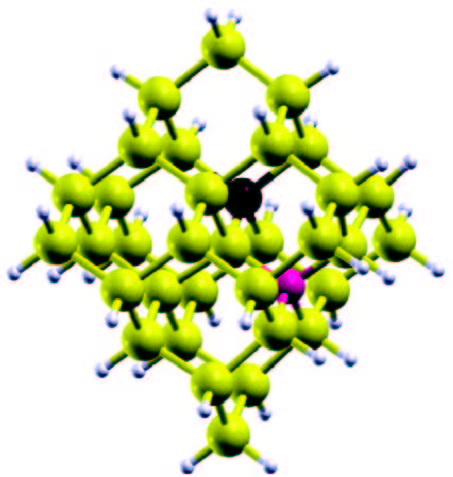}
\caption{\label{Si33BP}(Color online) Relaxed structure of the
Si$_{33}$BPH$_{36}$ codoped nanocrystal(d = 1.10 nm). Grey
(yellow) balls represent Si atoms, while the white (small grey)
balls are the hydrogens used to saturate the dangling bonds. B
(magenta, dark grey) and P (black) impurities have been located at
subsurface position in substitutional sites on opposite sides of
the nanocrystals. The relaxed impurity-impurity distance is
D$_{\text{BP}}$ = 3.64 {\AA}}
\end{figure}

\begin{figure}[!b]
\includegraphics[clip,width=0.35\textwidth]{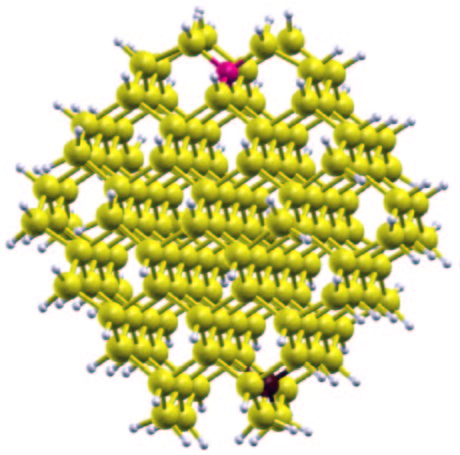}
\caption{\label{Si145BP}(Color online) Relaxed structure of the
Si$_{145}$BPH$_{100}$ codoped nanocrystal (diameter d = 1.79 nm).
Yellow (grey) balls represent Si atoms, while the white (small
grey) balls are the hydrogens used to saturate the dangling bonds.
B (magenta, dark grey) and P (black) impurities have been located
at subsurface position in substitutional sites on opposite sides
of the nanocrystals. The relaxed impurity-impurity distance is
D$_{\text{BP}}$ = 13.59 {\AA}}
\end{figure}

This tendency towards a $T_{d}$ symmetry of codoped Si-nc is also
obtained for smaller and larger nanocrystals, showing that these
outcomes are independent of the Si-nc size. Anyway, a symmetry
lowering with respect to Si$_{87}$H$_{76}$ is present due to the
different neighborhood experienced by the impurities. It should,
in fact, be noted that in the case of Si$_{35}$H$_{36}$ and
Si$_{147}$H$_{100}$ the atoms in the first subsurface shell are
bonded to three surface Si atoms and to one inner Si atom, while
for Si$_{87}$H$_{76}$ they are bonded to two surface and to two
inner Si atoms. The impurity positions for the considered
nanocrystals are showed in Fig. \ref {si85bp}, Fig. \ref{Si33BP}
and Fig. \ref{Si145BP}.
\begin{table}
\caption{\label{tableSi147}Bond lengths (in {\AA}) around impurity
sites for the undoped, single and codoped Si$_{147}$H$_{100}$
nanocrystal (d = 1.79 nm). Substitutional B and P impurities are
located at subsurface positions (see Fig. \ref{Si145BP}). Si$_s$
and Si$_i$ have the same meaning as in Table \ref{tableSi35}.}
\begin{ruledtabular}
\begin{tabular}{lccccc}
 & {Si$_{147}$H$_{100}$} & {}&{Si$_{146}$BH$_{100}$} & {Si$_{146}$PH$_{100}$} & {Si$_{145}$BPH$_{100}$} \\
{Bond} & {\AA} & Bond & {\AA} & {\AA} & {\AA} \\
\hline
Si-Si$_s$ & 2.356 &   B-Si$_s$     & 2.029     & & 2.016  \\
Si-Si$_s$ & 2.356   &  B-Si$_s$     & 2.029     & & 2.016\\
Si-Si$_s$ & 2.356   &  B-Si$_s$     & 2.063    & & 2.018\\
Si-Si$_i$ & 2.369   &  B-Si$_i$     & 2.009   & & 2.022\\
\hline
Si-Si$_s$ & 2.356   &  P-Si$_s$     &   &   2.310   & 2.306\\
Si-Si$_s$ & 2.356  &   P-Si$_s$     &   &   2.310   & 2.306\\
Si-Si$_s$ & 2.356   &  P-Si$_s$     &   &   2.372   & 2.338\\
Si-Si$_i$ & 2.369   &  P-Si$_i$     &   &   2.321   & 2.321\\
\end{tabular}
\end{ruledtabular}
\end{table}

Table II and III give the structural modifications that occur
around the impurities for Si$_{35}$H$_{36}$ and
Si$_{147}$H$_{100}$, respectively. Even in these cases the
differences between the four Si-impurity bond lengths on going
from the undoped to the single-doped to the codoped case first
increase and then decrease.

\begin{figure}
\includegraphics[clip,width=0.35\textwidth]{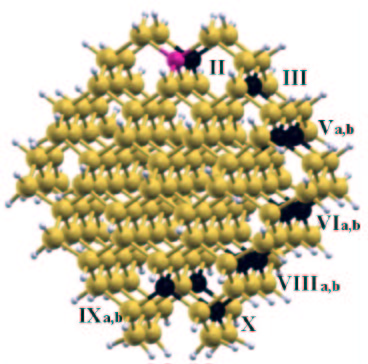}
\caption{\label{Si145BP_path}(Color online) ``Phosphorus impurity
path'' in Si$_{145}$BPH$_{100}$. Atoms have the same color as in
Fig. \ref{Si33BP}. The P atom (black) has been moved to explore
several substitutional sites (labeled by roman numbers) from the
position labeled by II to the positions III, V-a, V-b, VI-a, VI-b,
VIII-a, VIII-b, IX-a, IX-b and X. The B atom (magenta, dark grey)
position is fixed.}
\end{figure}

\begin{figure}[!b]
\includegraphics[clip,width=0.35\textwidth]{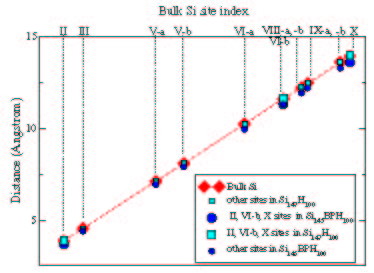}
\caption{\label{dist_dist}(Color online) B-P substitutional sites
distances in Si$_{145}$BPH$_{100}$ (blue circles) and the
corresponding Si-Si distances in the undoped Si$_{147}$H$_{100}$
nanocrystal (cyan squares) compared to the corresponding distances
in bulk Si (black diamonds). }
\end{figure}

Once stated that the amount of structural deformation remains
unvaried as the nanocrystal size changes and having shown that
this behavior is simply related to the codoping, we have devoted
our attention on how the presence of both the impurities acts. We
have looked, in particular, at what happens to the
impurity-impurity distance when compared to the corresponding
Si-Si distance in the undoped nanocrystal.
We have calculated these distances for Si$_{145}$BPH$_{100}$
keeping the B atom fixed in a subsurface position and moving the P
atom through different substitutional sites along the first
subsurface shell, as schematically shown in Fig.
\ref{Si145BP_path}.
We have moved the P atom from the position labeled II to the
positions III, V-a, V-b, VI-a, VI-b, VIII-a, VIII-b, IX-a, IX-b
and X. Here roman number refers simply to the positions evidenced
in Fig. \ref{Si145BP_path}. For each configuration, we have
calculated the B-P distance after a geometry relaxation and
repeated the calculation for the corresponding Si-Si distance. The
results are shown in Fig. \ref{dist_dist} where, as a reference,
we also show the corresponding distances in bulk silicon.
Taking bulk silicon as a reference, values above the dashed line
reflect an increase whereas values below correspond to a reduction
of the distances. It is seen that in all the cases considered the
distances are only weakly modified. Indeed, on going from the
undoped nanocrystals (where the distances are almost the same as
in bulk silicon) to the codoped ones we note a very small
shrinkage of the impurity-impurity distances. This shows, once
again, that if carriers are perfectly compensated by simultaneous
doping, the Si-nc does not really undergoes a significant
structural distortion, and this fact does not depend on the
distance between the impurities.

\subsection{Formation energy}
\label{forme} The different structural deformations occurring in
the single-doped and codoped nanocrystals around the impurity (see
Tables I, II and III) have a deep influence on the stability of the
analyzed systems. As stated in Sect. \ref{comp}, starting
from Si$_n$H$_m$,\cite{degoli} the formation energy of the
neutral B or/and P impurities can be defined as the energy needed
to insert one B or/and one P atom within the nanocrystal after
removing one/two Si atoms.

In order to clarify which are the parameters that play an important
role in the determination of the formation energy, we have
performed a series of total energy calculations considering: i)
single-doped and codoped nanocrystals, ii) nanocrystals of
different sizes, iii) impurities located in different sites and
iv) variable impurity-impurity distance within a nanocrystal.

\begin{figure}
\includegraphics[clip,width=0.35\textwidth]{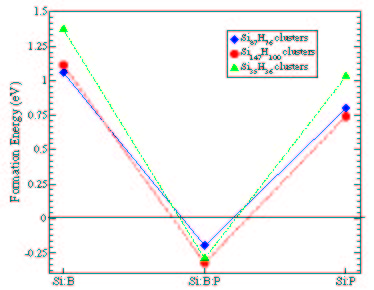}
\caption{\label{FormEnimpvicine}(Color online) Formation energy
for single-doped and codoped Si-nc. In the codoped nanocrystals
the impurities are placed as second neighbors in the first
subsurface shell (see text). Green triangles are related to
Si$_{35}$H$_{36}$, blue diamonds to Si$_{87}$H$_{76}$ and red
circles to Si$_{147}$H$_{100}$ based nanocrystals. The lines are a
guide for eyes. }
\end{figure}

In Fig. \ref{FormEnimpvicine} we report the calculated formation
energies of Si$_{35}$H$_{36}$ (diameter d= 1.10 nm),
Si$_{87}$H$_{76}$ (d= 1.50 nm) and Si$_{147}$H$_{100}$ (d = 1.79
nm).
In the same figure, as a reference, we report also the
single-doping formation energies. For the codoped case, B and P
impurities have been placed as second neighbors. This choice
corresponds to the nearest possible distance between two
subsurface sites for both Si$_{33}$BPH$_{36}$ (see Fig.
\ref{Si33BP}) and Si$_{145}$BPH$_{100}$ (see the position labeled
II in Fig. \ref{Si145BP_path})). After a geometry relaxation, the
distances between B and P impurities are D$_{\text{BP}}$ = 3.56
{\AA}, D$_{\text{BP}}$ = 3.64 {\AA}~ and D$_{\text{BP}}$ = 3.68
{\AA}~ for Si$_{33}$BPH$_{36}$, Si$_{85}$BPH$_{76}$ and
Si$_{145}$BPH$_{100}$ respectively.

From Fig. \ref{FormEnimpvicine} it is clear that the simultaneous B
and P doping strongly reduces (of about 1 eV) the formation energy
with respect to both B and P single-doped cases and that this
reduction is similar for Si-nc of different sizes. Thus, while B or
P single doping is very costly (in particular, the formation
energy increases with decreasing the nanocrystals size, in agreement
with previous calculations \cite{melnikov,cantele}), the
codoping is much easier and, as a good approximation, independent
of the nanocrystal size. The important point here is that Si-nc
can be more easily simultaneously doped than single-doped; this is
due to both the charge compensation and to the minor structural
deformation.

\begin{figure}
\includegraphics[clip,width=0.35\textwidth]{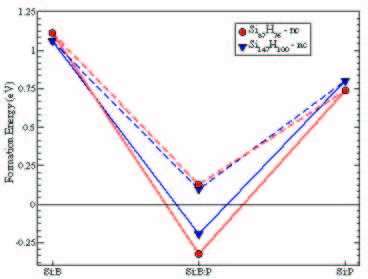}
\caption{\label{FormEn_confronto}(Color online)  Formation energy
of single-doped and codoped Si$_{87}$H$_{76}$ and
Si$_{147}$H$_{100}$ nanocrystals. Two different impurity-impurity
distances are considered in the codoped nanocrystals(dashed and
solid lines, larger and smaller distance respectively, see text).
Red circles refer to Si$_{87}$H$_{76}$, blue squares to
Si$_{147}$H$_{100}$. The lines are a guide for eyes.}
\end{figure}

It is interesting to look at the detailed dependence of the
formation energy on the distance between the two impurities. In
Fig. \ref{FormEn_confronto} we present the comparison between the
formation energies of Si$_{85}$H$_{76}$ and Si$_{145}$H$_{100}$
with impurities placed at two different distances : 1) the
previous considered second neighbors ones and 2) the largest
possible impurities distance (D$_{\text{BP}}$ = 10.60 {\AA}~ and
D$_{\text{BP}}$ = 13.29 {\AA}~ for the Si$_{85}$BPH$_{76}$ and
Si$_{145}$BPH$_{100}$ respectively, see Fig. \ref{si85bp} and Fig.
\ref{Si145BP}).
We note that when the impurity-impurity distance is reduced, the
formation energy decreases of 0.2-0.3 eV taking negative values.
This fact demonstrates that a stronger interaction between
impurities leads to a reduction in the formation energy, so that
codoping result to be easier and more likely when the dopants are
closer to each other. In the latter case, the reduction of the
formation energy is almost independent of the nanocrystal size, as
shown in Fig. \ref{FormEnimpvicine}.

\begin{figure}
\includegraphics[clip,width=0.35\textwidth]{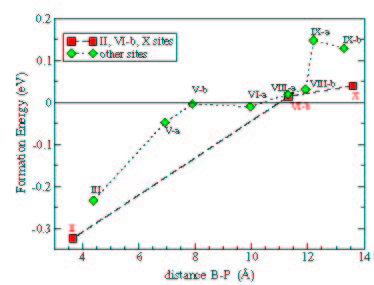}
\caption{\label{FormEn_path}(Color online) Formation energy as a
function of the boron-phosphorus distance. Roman numbers label the
positions of the P atom (see Fig. \ref{Si145BP_path}). The dotted
and dashed lines connect two subsets of impurity sites in which
the surrounding surface Si atoms are bounded to the same number of
passivating H atoms.}
\end{figure}

In order to investigate in more detail the dependence of the
formation energy on the impurity-impurity distance, we focus our
attention on the codoped Si$_{145}$BPH$_{100}$, trying to trace a
``formation energy path'' by progressively increasing the B-P
distance. In this calculation we have kept the B atom frozen in a
subsurface position while moving the P atom through different
substitutional sites along the first subsurface shell, as
schematically sketched in Fig. \ref{Si145BP_path}. The results of
these calculations are shown in Fig. \ref{FormEn_path}.

Two interesting effects are evidenced in this figure. The first
one is that the formation energy assumes a negative value when the
impurities are placed at distances smaller than 10 \AA, evolving
towards positive values for larger distances. This change of sign
can lead to the definition of a ``critical impurity distance''.
Below such a threshold the interaction between boron and
phosphorus is strong and gives rise to a reduction of the
formation energy. On the contrary, above this value, the
interaction tends to be quenched reducing the stability of the
impurity complex.

\begin{figure}[!b]
\includegraphics[clip,width=0.35\textwidth]{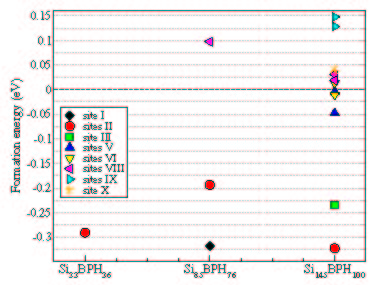}
\caption{\label{FormEn_cluster}(Color online) Formation energy as
a function of Si-nc size and impurity-impurity distance. For the
three different Si-nc one has the following nanocrystal diameters
(d) and impurity-impurity distances (D$_{\text{BP}}$):
Si$_{33}$BPH$_{36}$ d= 1.10 nm and D$_{\text{BP}}$ = 3.56 {\AA}
(site II); Si$_{85}$BPH$_{76}$ d= 1.50 nm and D$_{\text{BP}}$ =
2.00 {\AA} (site I), D$_{\text{BP}}$ = 3.64 {\AA} (site II),
D$_{\text{BP}}$ = 10.60 {\AA} (site VIII); Si$_{145}$BPH$_{100}$
d= 1.79 nm and D$_{\text{BP}}$ = 3.68 {\AA} (site II),
D$_{\text{BP}}$ = 4.40 {\AA} (site III), D$_{\text{BP}}$ = 6.93
{\AA} (site V-a), D$_{\text{BP}}$ = 7.91 {\AA} (site V-b),
D$_{\text{BP}}$ = 9.95 {\AA} (site VI-a), D$_{\text{BP}}$ = 11.32
{\AA} (site VI-b), D$_{\text{BP}}$ = 11.30 {\AA} (site VIII-a),
D$_{\text{BP}}$ = 11.93 {\AA} (site VIII-b), D$_{\text{BP}}$ =
12.20 {\AA} (site IX-a),D$_{\text{BP}}$ = 13.29 {\AA} (site IX-b),
D$_{\text{BP}}$ = 13.59 {\AA} (site X).}
\end{figure}

These considerations are also supported by Fig.
\ref{FormEn_cluster} where we report the values of the formation
energy for three different nanocrystals in which the impurities
are always located in the subsurface shells at different
distances.
As before, it is evident from this figure that the distance
between impurities plays a fundamental role on the decrease of the
formation energy. For each nanocrystal, the formation energy takes
on negative values below a given distance. Moreover, the formation
energy have a minimum value when the impurities are located at the
minimum possible distance. Indeed, the impurity-impurity distance
seems to play a major role with respect to the nanocrystals size,
since the formation energy for similar impurity configurations are
quite independent of the nanocrystal dimension. The small
difference between the Si$_{85}$BPH$_{76}$ and the
Si$_{33}$BPH$_{36}$ and Si$_{145}$BPH$_{100}$ is due to the
different neighborhood experienced by the impurities in the three
cases (see Tables I, II and III).

Another relevant point is the possibility to identify two
distinct trends for the formation energy (see Fig.
\ref{FormEn_path}) that can be related to the type of silicon cage
surrounding the P dopant site. One can group (dotted line)
together the cases in which the P impurity is located in the
positions labeled II, VI-b and X with respect to the B impurity
(see Fig. \ref{Si145BP_path}). In these positions two of the
surface Si atoms bounded to the P impurity present two passivating
H atoms instead of one, a situation that dominates in
all the other configurations. A different number of capping H atoms
influences the formation energy.

\subsection{Electronic Properties}
\label{elpr}

In this section we will investigate the role of codoping on the
electronic properties of Si-nc. As in the corresponding bulk
system, the insertion of impurities tends to modify the electronic
structure. We shall show that, by properly controlling the doping
and the size, it is possible to modulate both the electronic
structure and some optical features. In particular, we shall show
how the electronic properties of the codoped nanocrystals depend
on both the nanocrystal size and on the distance between the
impurities. Some of the results will be discussed in terms of wave
function localization around the impurity complex.

In the single-doped cases we have already shown that the presence
of either donor or acceptor states can considerably lower the
energy gap (E$_G$, the HOMO-LUMO energy difference) of the undoped
Si-nc,\cite{cantele,feffe} defining in this case the energy gap as
the gap between the impurity level (partially filled considered as
the HOMO) and the LUMO (which is empty). In these cases the
partially filled HOMO level is strongly localized either on the B
or on the P impurity. For example, in the case of
Si$_{86}$BH$_{76}$ the defect level is located just 0.28 eV above
the valence band reducing the above defined energy gap from 2.59
eV (the value for the undoped Si-nc) to 2.31 eV. In
Si$_{86}$PH$_{76}$ the defect level is located just 0.28 eV below
the conduction band so that the energy gap is only 0.28
eV.\cite{feffe} It is interesting to note that the experimental
substitutional donor binding energy for P in bulk Si is about 33
meV, while the experimental acceptor energy for B in Si is 45
meV,\cite{cardona} showing how, in the case of nanocrystals, the
combined effects of both quantum confinement and weak screening
tend to ``transform'' shallow impurities in ``deep''
centers.\cite{melnikov,cantele,ninno,tra2}

\begin{figure}
\includegraphics[clip,width=0.35\textwidth]{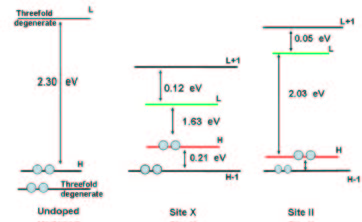}
\caption{\label{Si145-liv}(Color online) Calculated energy levels
of the undoped Si$_{147}$H$_{100}$ (left panel), codoped
Si$_{145}$BPH$_{100}$  with an impurity-impurity distance
D$_{\text{BP}}$ = 13.29 {\AA} (central panel), codoped
Si$_{145}$BPH$_{100}$ with an impurity-impurity distance
D$_{\text{BP}}$ = 3.68 {\AA} (right panel). The alignment has been
done locating at the same energy the fully occupied levels with
the same type of localization. Site X and Site II are referred to
Fig. \ref{Si145BP_path}. H stands for HOMO, L for LUMO.}
\end{figure}

The electronic properties of B- and P- codoped Si-nc are
qualitatively and quantitatively different from those of either B-
or P- single-doped Si-nc. Now the system is a semiconductor and
the presence of both the impurities leads to a HOMO level that
contains two electrons and to a HOMO-LUMO energy gap strongly
lowered with respect to that of the corresponding undoped
nanocrystals. Fig. \ref{Si145-liv} shows the energy levels of
Si$_{147}$H$_{100}$ and Si$_{145}$BPH$_{100}$ with the impurities
located at two different distances.

In one case, the impurities are placed at the largest possible
distance (D$_{\text{BP}}$ = 13.29 {\AA}) and in the other one at
the already discussed minimum distance (D$_{\text{BP}}$ = 3.68
{\AA}) for this particular nanocrystal. From the figure it is
evident that when impurities are at the larger distance, E$_G$ is
strongly reduced with respect to the corresponding undoped value
(E$_G$ is lowered from 2.30 eV to 1.63 eV). On the contrary, when
the impurities are close to each other, E$_G$ enlarges (E$_G$ =
2.03 eV) although it still remains below the undoped case. We can
think that when impurities are brought closer, the Coulomb
interaction becomes stronger so that the energy gap becomes
larger. Boron and phosphorus feel each other like a B-P complexes
with a gap opening recalling the DFT-LDA calculated gaps of the
boron phosphide bulk system: direct gap
$(\,\Gamma\,\rightarrow\Gamma\,)$ 3.3 eV, indirect gaps
(\,$\Gamma\,\rightarrow\,X$\,) 2.2 eV, and
(\,$\Gamma\,\rightarrow\,\Delta$\,) 1.2 eV, as described in Ref.
\onlinecite{bp}. \noindent

\begin{figure}
\includegraphics[clip,width=0.35\textwidth]{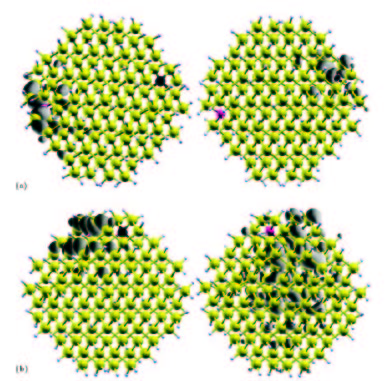}
\caption{\label{Si145-plots}(Color online) From the top to the
bottom: the HOMO (left) and LUMO (right) square modulus contour
plots calculated for Si$_{145}$BPH$_{100}$ (atom colors same as in
Fig. \ref{Si33BP}. The impurities are located on opposite sides of
the nanocrystal, at distance D$_{\text{BP}}$ = 13.29 {\AA} (a) or
as second neighbors, with an impurity-impurity distance
D$_{\text{BP}}$ = 3.68 {\AA}. The isosurfaces correspond to 10\%
of the maximum value.}
\end{figure}

These behaviors are corroborated by the calculated HOMO and LUMO
wave functions. Fig. \ref{Si145-plots} shows the square modulus
contour plots of the HOMO and LUMO states of the two considered
Si$_{145}$BPH$_{100}$ nanocrystals.

The top panel shows the contour when the impurities are at a large
distance while the bottom panel is that with the impurities at
short distance. It clearly appears from these contours that on
going from the case with well separated impurities to the that
with close impurities, the overlap between the HOMO, strongly
centered on the boron atom, and the LUMO, mainly localized on the
phosphorus atom, strongly increases.

\begin{figure}[!b]
\includegraphics[clip,width=0.35\textwidth]{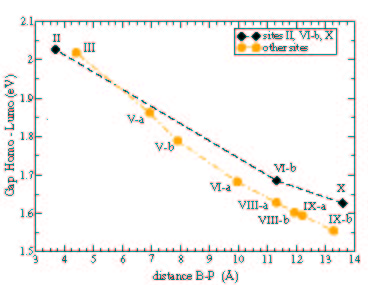}
\caption{\label{hlgap}(Color online) The HOMO-LUMO energy gap as a
function of the distance between the B and P impurities within the
Si$_{145}$BPH$_{100}$ nc. Roman numbers refer to the positions of
the P atom (see Fig. \ref{Si145BP_path}). The dashed and
dash-dotted lines connect the two subsets of impurity sites in
which the surrounding surface Si atoms are bonded to the same
number of passivating H atoms.}
\end{figure}

Next we investigate how the electronic structure changes as a
function of the impurity distance within the Si$_{145}$BPH$_{100}$
nanocrystal. In Fig. \ref{hlgap} we report the trend of the
HOMO-LUMO energy gap with respect to the distance between
impurities.
It is seen that the mutual impurity distance affects not only the
formation energy (see Sec. \ref{forme}), but also the electronic
structure. We observe that  E$_G$ decreases almost linearly with
the increase of the impurity distance; moreover, also in this case
we can figure out the presence of two different trends related to
the different surface region experienced by the P atom in the
sites II, VI-b and X, with respect to the other ones (see also
Fig. \ref{FormEn_path} and related discussion). Fig. \ref{hlgap}
points out how, at least in principle, it is possible to tune
E$_G$ as a function of the impurity-impurity distance. It is easy
to predict that for Si-nc larger than those considered here it
would be possible by codoping to obtain a energy gap even smaller
than that of bulk Si.

\begin{figure}
\includegraphics[clip,width=0.35\textwidth]{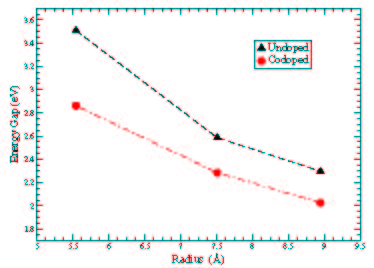}
\caption{ \label{Sinc-gaps}(Color online) Comparison between
energy gaps of the undoped (black triangles) and the codoped (red
circles) nanocrystals as a function of the nanocrystal radius.
Impurities are located in the first shell below the surface, as
second neighbors. The impurity-impurity distances are 3.56 \AA,
3.64 \AA, and 3.68 \AA~ for Si$_{33}$BPH$_{36}$,
Si$_{85}$BPH$_{76}$, Si$_{145}$BPH$_{100}$ respectively. The lines
are a guide for the eye.}
\end{figure}

The possibility of modulating the electronic properties of the
codoped Si-nc is also evident if we keep the distance between the
impurities constant and look at the dependence of the energy gap
on the Si-nc size. Fig. \ref{Sinc-gaps} shows, for three different
nanocrystals where the impurities are placed as second neighbors,
how the undoped nanocrystal energy gap is reduced in the presence
of codoping (see also Table IV).

The same quantum confinement effect trend (i.e. larger gap for
smaller nanocrystals) is observed for both the undoped and the
codoped cases. Moreover, the energy gap of the codoped Si-nc is
shifted towards lower energies with respect to that of the undoped
E$_G$; this shift is stronger for smaller nanocrystals. Playing
with both the nanocrystal size and the distance between the
impurities new interesting routes may be opened for optoelectronic
applications.
\begin{table}[!ht]
\caption{\label{tableSigap} HOMO-LUMO gap ($E_G$) for
Si$_{35}$H$_{36}$, Si$_{87}$H$_{76}$, Si$_{147}$H$_{100}$ and the
corresponding codoped Si$_{33}$BPH$_{36}$, Si$_{85}$BPH$_{76}$,
and Si$_{145}$BPH$_{100}$. Impurities are second neighbors. The
impurity-impurity distances are 3.56 \AA, 3.64 \AA, and 3.68 \AA,
for Si$_{33}$BPH$_{36}$, Si$_{85}$BPH$_{76}$, and
Si$_{145}$BPH$_{100}$ respectively. d is the nanocrystal
diameter.}
\begin{ruledtabular}
\begin{tabular}{lccc}
{starting nc} & {{{d} (nm)}} & {\small{E$_G$ undoped (eV)}} & {\small{E$_G$ codoped (eV)}}\\
\hline
Si$_{35}$H$_{36}$    &1.10 & 3.51    & 2.86 \\
Si$_{87}$H$_{76}$    &1.50 & 2.59    & 2.29 \\
Si$_{147}$H$_{100}$  &1.79 & 2.30    & 2.03 \\
\end{tabular}
\end{ruledtabular}
\end{table}
Looking at the energy gap trends in Fig. \ref{hlgap} and Fig.
\ref{Sinc-gaps} and considering that in the codoped case Fujii et
al. \cite{fujii1} found photoluminescence peaks centered in the
0.9-1.3 eV energy region, we may conclude that Si-nc playing a
role in the experiment have dimensions of the order of few
nanometers. This conclusion is consistent with the experimental
outcomes \cite{fujii1} that indicates an average nanocrystal
diameter of about 5 nm.

\subsection{Optical Properties}
\label{opti} The aim of this section is to investigate the
mechanisms involved in the modification of the optical properties
of codoped Si nanocrystals. We present absorption and emission
spectra with a comparison between the IP-RPA (independent
particle-random phase approximation) spectra and the many-body
ones. These last ones are obtained within a GW-BSE approach that
takes into account not only the self-energy correction and the
local field effects but also the electron-hole interaction. All
the calculations performed are not spin-polarized. However it
should be noted that single-particle calculations for undoped
Si-nc have been done by Franceschetti and Pantelides
\cite{franceschetti} within the local spin-density approximation,
showing that the singlet-triplet splitting is significantly
smaller than the Stokes shift. To understand the role of
dimensionality and impurity distance and to show the importance of
including many-body effects in the optical spectra, we are going
to present first the result of a RPA independent particle optical
response for various codoped nanocrystals different in dimensions
and in impurity location (see Sect. \ref{abs}), and next, we will
present a complete study of a codoped Si-nc where we go beyond the
single-particle approach within the GW-BSE framework (see Sect.
\ref{lum}).

\subsubsection{Absorption and emission spectra: single-particle results}
\label{abs}

\begin{figure}
\includegraphics[clip,width=0.35\textwidth]{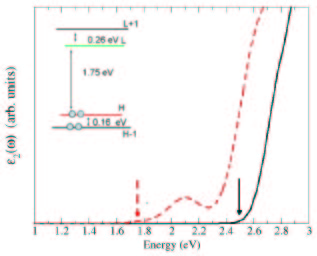}
\caption{\label{Si85BP_abs}(Color online) Comparison between the
undoped Si$_{87}$H$_{76}$ (solid line) and the codoped
Si$_{85}$BPH$_{76}$ (dashed line) single-particle absorption
spectra. The impurities are at a distance of 10.60 \AA. Arrows
indicate the energy gaps. The calculated energy levels for the
codoped nanocrystal are shown in the inset. A Gaussian broadening
of 0.1 eV has been applied. H stands for HOMO, L for LUMO. }
\end{figure}

We first discuss the results related to the absorption spectra.
Fig.\ref{Si85BP_abs} shows a comparison between the undoped
Si$_{87}$H$_{76}$ and the codoped Si$_{85}$BPH$_{76}$ IP-RPA
absorption spectra.
In this case the impurities are located at a distance of 10.60
\AA. The optical response is evaluated for the ground state
relaxed geometry computing the imaginary part of the dielectric
function $\epsilon_2(\omega)$. It can be seen from
Fig.\ref{Si85BP_abs} that new transitions arise below the
absorption onset of the undoped Si-nc. In particular, we have
found a shift of the absorption gap to lower energies with respect
to the undoped case toghether with an enhancement of the
intensities around 2.0 eV. These new transitions are due to the
presence of new HOMO and LUMO states localized on the impurities,
as described in Sec. \ref{elpr} (see for example Fig.
\ref{Si145-plots}). The inset of Fig. \ref{Si85BP_abs} clarifies
how the peak located in the 2.0-2.2 eV energy region is related to
contributions that involve the HOMO-1, HOMO to LUMO, LUMO + 1
transitions; it should be noted that for all these levels the
wavefunctions are predominantly localized on the impurities. If we
compare these results with those of a single-doped
Si-nc,\cite{feffelumin} we note that the simultaneous presence of
both impurities naturally suppresses all the absorption energy
structures present in the infrared region (below 1 eV) of the
single-doped spectra.
\begin{figure}[!b]
\includegraphics[clip,width=0.35\textwidth]{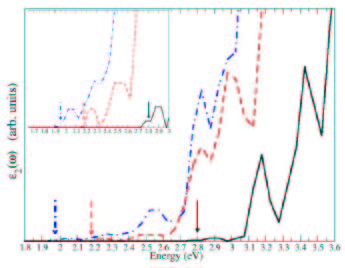}
\caption{\label{Siabsdim}(Color online) Single-particle absorption
spectra of Si$_{33}$BPH$_{36}$ (black solid line),
Si$_{85}$BPH$_{76}$ (red dashed line), and Si$_{145}$BPH$_{100}$
(dash-dotted blue line). In all cases the impurities are second
neighbors. The impurity-impurity distances are 3.56 \AA, 3.64 \AA,
and 3.68 \AA~ for Si$_{33}$BPH$_{36}$, Si$_{85}$BPH$_{76}$,
Si$_{145}$BPH$_{100}$ respectively. Arrows indicate the energy
gaps. In the inset a zoomed view of the spectra onset. No Gaussian
broadening has been applied. }
\end{figure}
It is clear that, like the electronic properties, also the optical
ones present a marked dependence on the nanocrystal dimension. To
elucidate this point we plot in Fig. \ref{Siabsdim} the
single-particle absorption spectra of three different Si-nc, the
Si$_{33}$BPH$_{36}$, Si$_{85}$BPH$_{76}$, Si$_{145}$BPH$_{100}$,
whose diameters are 1.10 nm, 1.50 nm and 1.79 nm, respectively. In
all the three nanocrystals the impurities are second neighbors.

Two facts emerge from this figure. First of all, on increasing the
nanocrystal size the absorption gap is strongly reduced (see
arrows in Fig. \ref{Siabsdim}). Second, an increase of the Si-nc
diameter (i.e. a decrease of the impurity weight with respect to
the total number of atoms) results in a lowering of the intensity
for the transitions that involve the impurities.

\begin{figure}
\includegraphics[clip,width=0.35\textwidth]{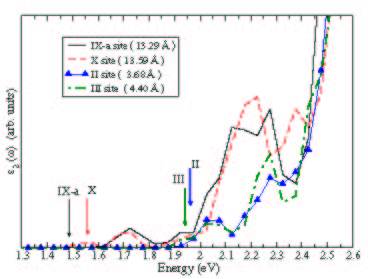}
\caption{\label{Si145BP-abs}(Color online) Single-particle
absorption spectra of Si$_{145}$BPH$_{100}$ Si-nc with impurities
placed at different distances (see the inset). Arrows indicate the
energy gaps. Roman numbers refer to the sites occupied by the P
atom with respect to the B one (see Fig. \ref{Si145BP_path}). No
Gaussian broadening has been applied. }
\end{figure}

The role of the impurity distance on the optical response
has been investigated following the same approach adopted for the
electronic properties in Sec. \ref{elpr}. In
Fig.\ref{Si145BP-abs} the single-particle absorption spectra of
Si$_{145}$BPH$_{100}$ are reported with P impurity
placed on sites II, III, IX, and X respectively (see for comparison
Fig. \ref{Si145BP_path}).

Here we observe a shift of the absorption gap to lower energy on
increasing the distance between the impurities (see arrows in
Fig.~\ref{Si145BP-abs}). Moreover, also the intensity is affected
by the impurity distance. Stronger transitions arise when the
impurities are closer whereas the intensity gets lower when the
impurities are at a larger distances; the optical transitions near
the band edge (indicated by arrows in Fig.~\ref{Si145BP-abs})
exhibit weaker oscillator strengths.

Now we discuss the results for the emission spectra and for the
Stokes shift between absorption and emission. The nanoscrystal
excitation has been studied considering the excited state as the
electronic configuration in which the highest occupied
single-particle state (HOMO) contains a hole ($h$), while the
lowest unoccupied single-particle state (LUMO) contains the
corresponding electron ($e$), thus simulating the creation of a
electron-hole pair.\cite{godby,frang,weisskerprl90,puzder}
Initially the system is in its ground state and the electronic
excitation occurs with the atomic positions fixed in this
configuration. After the excitation, due to the change in the
charge density, relaxation occurs until the atoms reach a new
minimum energy due to the presence of the electron-hole pair. The
new atomic positions modify the electronic spectrum, implying that
the levels involved in the emission process change. This model
assumes that the relaxation under excitation is faster than the
electron-hole recombination. The difference between the absorption
and emission energies due to the different atomic positions
represents the nanocrystal Stokes
shift.\cite{franceschetti,leo,comp}

The calculations have been performed for two Si-nc of different
sizes taking, in the larger Si-nc, the impurities located at
different distances. As shown in Table V, both the absorption and
emission HOMO-LUMO energies are affected by these two parameters.
\begin{table}
\caption{\label{table-gaps-DFT} Absorption and emission energy
gaps (and their difference, 5th row) calculated as HOMO-LUMO
differences in the ground and the excited relaxed geometries
configuration, respectively. The results are obtained within the
DFT single-particle approach. d is the nanocrystal diameter,
D$_{\text{BP}}$ is the distance between impurities and $\Delta$
the calculated Stokes shift between absorption and emission energy
gaps.}
\begin{ruledtabular}
\begin{tabular}{lccc}
& Si$_{33}$BPH$_{36}$ &  Si$_{85}$BPH$_{76}$ & \\
\hline
d (nm) & 1.10 & 1.50  & 1.50 \\
D$_{\text{BP}}$ (\AA) & 3.56 & 2.00  & 10.60 \\
Abs. (eV)     &    2.77 & 2.32  & 1.75 \\
Ems. (eV)    &    1.78 & 2.20 & 1.36\\
$\Delta$\,\,(eV) & 0.99 & 0.12 & 0.39
\end{tabular}
\end{ruledtabular}
\end{table}
With regard to the first parameter, we note that the Stokes shift
strongly depends on the size showing a strong decrease on
increasing the diameter of the Si-nc. This is due to the fact that
for larger nanocrystals the excitation determines a minor
distortion of the geometry. Concerning the second parameter, we
see that the Stokes shift tends to slightly increase with B-P
distance although this effect is small if compared with the
lowering due to the increase of the Si-nc dimensions. The
comparison between these results and the ones previously obtained
for undoped clusters (0.92 eV for the
Si$_{35}$H$_{36}$-nc\cite{degoli} and 0.22 eV for the
Si$_{87}$H$_{76}$-nc\cite{puzder}) confirm that the Stokes shifts
is mainly determined by the nanocrystals size, but that
nevertheless it depend slightly on the presence of the impurities
and also on their mutual distance.

\begin{figure}
\includegraphics[clip,width=0.35\textwidth]{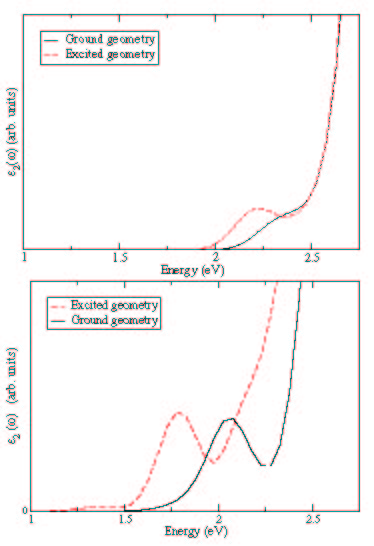}
\caption{\label{Si85BP-abs-em}(Color online) Single-particle
imaginary part of the dielectric function for the codoped
Si$_{85}$BPH$_{76}$ nanocrystal in the ground (black-solid line)
and in the excited (red-dashed line) geometries. B and P atoms are
at the smallest possible distance (2.00 \AA, top panel) or at the
largest possible distance (10.60 \AA, bottom panel) for this
nanocrystal. A Gaussian broadening of 0.1 eV has been applied.}
\end{figure}

Looking at the single-particle optical spectra in Fig.
\ref{Si85BP-abs-em} we note that the HOMO-LUMO transition in
Si$_{85}$BPH$_{76}$ (1.75 eV, bottom panel) is almost dark when
the two impurities are far apart and becomes instead allowed (2.32
eV, top panel) when their distance decreases. As discussed before,
this oscillator strength enhancement is a consequence of the
character of the HOMO and LUMO states in the two cases.
The emission (red-dashed lines in Fig. \ref{Si85BP-abs-em})
spectra is red shifted with respect to the absorption (black-solid
lines in Fig. \ref{Si85BP-abs-em}). This red shift is a
consequence of the geometry relaxation in the excited state due to
the excess energy necessary for promoting of an electron in the
LUMO level. The dependence of the emission spectra both on the
nanocrystals size (see Table V and Fig. \ref{Siabsdim}) and on the
impurities positions (see Figs. \ref{Si145BP-abs} and
\ref{Si85BP-abs-em}) reveals once more the possibility of tuning
the optical response of silicon nanocrystals.

\subsubsection{Absorption and Emission Spectra: Many-body effects}
\label{lum}

In order to give a complete description, within the many-body
framework, of the codoped Si-nc response to an optical excitation,
we consider both the self-energy corrections by means of the GW
method \cite{GWdetails} to obtain the quasiparticle energies and the
excitonic effects through the solution of the Bethe-Salpeter
equation.  The effect of local fields is also included, to take into
account the inhomogeneity of the systems.

\begin{figure}
\includegraphics[clip,width=0.35\textwidth]{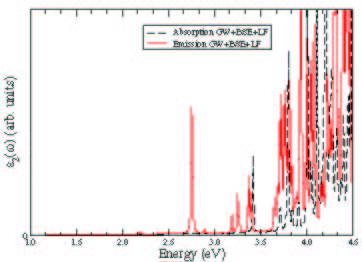}
\caption{\label{Si33abs-ems}(Color online) Absorption
(black-dashed line) and emission (red -solid line) many-body
spectra of Si$_{33}$BPH$_{36}$.}
\end{figure}

Since the GW-BSE calculation \cite{exccode} are very computing
demanding, we have only considered the smaller codoped nanocrystal
Si$_{33}$BPH$_ {36}$ (see Fig. \ref{Si33BP}). In this particular
cluster, we found that Local fields effects are, although not
negligible, of minor importance with respect to GW and excitonic
effects. It is anyway essential to include {\it all} of them (LF
and many-body) in order to get the final converged spectrum shown
in Fig.  \ref{Si33abs-ems}.

In order to carry out emission spectra calculations, we use the
excited state geometry and the ground state electronic
configuration. As already noted before, in this case
$\epsilon_2(\omega)$ corresponds to an absorption spectrum in a
new structural geometry. In other words, we consider the emission
as the time reversal of the absorption.\cite{bassani,leo} Thus,
the electron-hole interaction is here considered also in the
emission geometry. The heavy GW-BSE calculation is made
considering a large FCC supercell with a 50 a.u. lattice
parameter.  The correlation part of the self-energy $\Sigma_c$ has
been calculated using 10081 plane waves, while 49805 plane waves
have been used for the exchange part $\Sigma_x$. Then, the full
excitonic Hamiltonian is diagonalized considering more than 8000
transitions.

Fig. \ref{Si33abs-ems} shows the calculated absorption and
emission spectra fully including the many-body effects. The
electron-hole interaction yields significant variations with
respect to the single-particle spectra (see for a comparison Fig.
\ref{Si85BP-abs-em}), with an important transfer of the oscillator
strength to the low energy side. Moreover, in the emission
spectrum the rich structure of states characterized, in the low
energy side, by the presence of excitons with largely different
oscillator strengths, determines excitonic gaps well below the
optical absorption onset. Thus the calculated emission spectrum
results to be red shifted to lower energy with respect to the
absorption one. This energy difference between emission and
absorption, the Stokes shift, can be lead back to the relaxation
of the Si-nc after the excitation process.

The new important features that appear in the emission many-body
spectra are related to the presence of both B and P impurities as
showed by Fig. 19, which gives the real-space probability
distribution $|\psi_{exc}(r_e,r_h)|^2$ for the bound exciton as a
function of the electron position $r_e$ when the hole is fixed in a
given $r_h$ position. In this case the hole is fixed on the boron
atom and we see that the bound exciton is mainly localized around
the phosphorus atom.
\begin{figure}
\includegraphics[clip,width=0.35\textwidth]{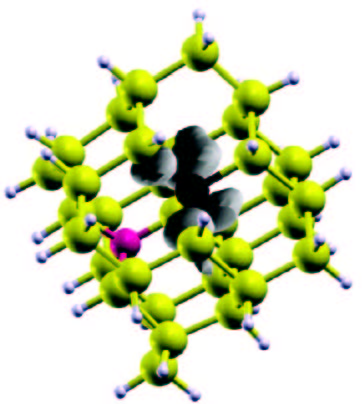}
\caption{\label{hhheee}(Color online) Excitonic wave function of
Si$_{33}$BPH$_{36}$ (atom colors as in Fig. \ref{Si33BP}). The
grey isosurface represents the probability distribution
$|\psi_{exc}(r_e,r_h)|^2$ of the electron with the hole fixed on
the B impurity. }
\end{figure}

From Table VI, it can be seen that the single-particle DFT results
strongly underestimate the absorption and emission edge with
respect to the GW+BSE calculation, in which the excitonic effect
are taken exactly into account. This means that, in this case, the
cancellation between GW gap opening (which gives the electronic
gap) and BSE gap shrinking (which originates the excitonic gap) is
only partial.\cite{del}

The difference between the GW electronic gap and the GW+BSE
optical excitonic gap gives the exciton binding energy E$_b$. We
note the presence of exciton binding energies as big as 2.2\,eV,
which are very large if compared with bulk Si ($\sim$ 15 meV) or
with carbon nanotubes\cite{spataru,chang} where E$_b\sim$\,1\,eV,
but similar to those calculated for undoped Si-nc \cite{leo} of
similar size and for Si and Ge small
nanowires.\cite{bruno1,bruno2}
\begin{table}[!hb]
\caption{\label{table-gaps} Absorption and Emission energies
calculated as HOMO-LUMO energy difference within the
single-particle DFT, the many-body GW and the GW+BSE approaches.
$\Delta$ is the calculated Stokes shift between absorption and
emission energy gap. The 2.20 eV energy corresponds to an almost
dark transition.}
\begin{ruledtabular}
\begin{tabular}{lccc}
Si$_{33}$BPH$_{36}$ & DFT  &  GW & GW+BSE  \\
\hline
Abs. (eV)     &    2.80 &  5.52 & 3.35 \\
Ems. (eV)    &    1.79 & 4.37 & 2.20\\
$\Delta$\,\,(eV) & 1.01 & 1.15 & 1.15
\end{tabular}
\end{ruledtabular}
\end{table}

The differences between full many-body calculations and
single-particle results are of 0.55 eV and 0.41 eV for absorption
and emission energy gaps respectively, and of 0.14 eV between the
two Stokes shifts. It is interesting to note that the HOMO-LUMO
transition in the emission spectrum at 2.20 eV is almost dark while
an important excitonic peak is evident at about 2.75 eV (see Fig.
\ref{Si33abs-ems}), again red-shifted with respect to the first
absorption peak. As expected, what comes out is the importance of
fully taking into account the many-body aspect of the problem in
order to overcome the limits of the single-particle approach.
\section{Conclusions}
\label{conc}

The structural, electronic and optical properties of Si
nanocrystals codoped with B and P impurities have been studied
also going beyond the single-particle approach. We have considered
Si-nc of different size and with the impurities located at
different distances.  We show that codoping is always
energetically favored with respect to simple B- or P-doping and
that the two impurities tend to occupy nearest neighbor
sites near the surface rather than other positions inside the
nanocrystal itself. Our results demonstrate that the codoped
nanocrystals present valence and conduction band-edge states which
are localized on the two impurities respectively and band-gaps
always lower in energy with respect to that of undoped Si
nanocrystals. Besides, the electronic properties show a dependence
on both nanocrystal size and impurity-impurity distance. The
impurity located band-edge states originate absorption thresholds
in the visible region which are shifted lower in energy with
respect to the undoped case. Moreover, the
emission spectra show a Stokes shift with respect to the
absorption which is due to the structural relaxation after the
creation of the electron-hole pair. Our results make evident the
presence of electronic quasi-direct optical transitions between
donor and acceptor states that can help to understand the
experimental outcomes and makes it possible to engineer the
absorption and emission properties of Si nanocrystals.

\section*{Acknowledgments}
We acknowledge the support of the MIUR PRIN (2005) Italy, of the
CNR-CNISM "Progetto Innesco"  of the CRUI Vigoni Project
(2005-2006) Italy-Germany and of  EU
Nanoquanta Network of Excellence (NMP4-CT-2004-500198).
 All the calculations were performed at
CINECA-Bologna ("Iniziativa Calcolo Parallelo del CNR-INFM"),
CICAIA-Modena and ``Campus Computational Grid''-Universit\`a di
Napoli ``Federico II''.


\begin{thebibliography}{}
\bibitem{check} We have checked that the use of Vanderbilt or norm-conserving pseudopotentials
does not affect the calculated structural properties of the Si-nc.
\bibitem{comp} We note that the difference that arises in calculating the energy gap using  the HOMO-LUMO gap or the $\Delta$-SCF
method strongly shrinks as the dimension of the nanocrystal
increases \cite{degoli,puzder}.
\bibitem{GWdetails} We have used the non selfconsistent
G$_{0}$W$_{0}$ approach within the RPA plasmon pole approximation.
We use a planewave-frequency space code.
\end{thebibliography}
\end{document}